\documentclass{iopconfser}

\usepackage{graphicx}

\begin{document}

\newcommand {\nc} {\newcommand}
\nc {\bce}{\begin{center}}
\nc {\ece} {\end{center}}
\nc {\bit} {\begin{itemize}}
\nc {\eit} {\end{itemize}}
\nc {\beq} {\begin{eqnarray}}
\nc {\eeq} {\end{eqnarray}}
\nc {\eeqn}[1] {\label {#1} \end{eqnarray}}
\nc {\flim} [2] {\mathop{\longrightarrow}\limits_{{#1}\rightarrow{#2}}}
\nc {\ve} [1] {\mbox{\boldmath $#1$}}
\nc {\etal} {\emph{et al.\ }}
\nc {\eq} [1] {(\ref{#1})}
\nc {\Eq} [1] {Eq.\,(\ref{#1})}
\nc {\Fig} [1] {Fig.\,\ref{#1}}
\nc {\Sec} [1] {Sec.\,\ref{#1}}

\title{Halo-EFT: an effective and efficient tool to study reactions with halo nuclei}

\author{Quentin Bozet$^{1}$, Live-Palm Kubushishi$^{2,1}$ and Pierre Capel$^{1}$}

\affil{$^1$Institut f\"ur Kernphysik, Johannes Gutenberg-Universit\"at, 55099 Mainz, Germany\\
$^2$Institute of Nuclear and Particle Physics and Department of Physics and Astronomy, Ohio University, Athens, OH
45701,USA}

\email{pcapel@uni-mainz.de}

\begin{abstract}
Halo nuclei are exotic nuclei, which exhibit a much larger matter radius than their isobars.
This unusual size is now understood as a threshold effect, in which one or two valence nucleons are loosely bound to the core of the nucleus.
Thanks to this loose binding, halo nuclei exhibit a strongly clusterised structure with a diffuse neutron halo surrounding a compact core.
Being short lived, halo nuclei are mostly studied through reactions, such as breakup.
In this talk we show that including Halo Effective Field Theory within existing reaction codes is both effective and efficient to analyse experimental data.
We first apply this idea to study the Coulomb breakup of $^{19}$C within a Bayesian approach.
Second, we extend Halo-EFT to include the core's excitation to test the sensitivity of the Coulomb breakup of $^{11}$Be on spectroscopic factors.
\end{abstract}

\section{Introduction}
The development of radioactive-ion beam facilities in the mid-80s, has enabled us to explore the nuclear chart far from stability.
This led to the discovery of the halo structure in light nuclei close to the neutron dripline \cite{Tan85b}.
These exotic nuclei exhibit a much larger matter radius than their isobars.
This exceptional size is due to their small separation energy for one or two neutrons.
The weak binding leads to an extended wave function for these valence neutrons, which thus exhibit a high probability of presence at large distance from the other nucleons.
Halo nuclei can thus be seen as a highly clusterised system with a diffuse and extended neutron halo surrounding a dense and compact core \cite{Tan96}.

Because of their short lifetime, halo nuclei are mostly studied through reactions, such as Coulomb breakup.
In that reaction, the halo nucleus collides with a heavy target, within which field it dissociates into its core and halo neutron.
To infer reliable structure information from such measurement, we need an accurate model of the reaction coupled to a realistic description of the projectile.
Halo-Effective Field Theory (Halo-EFT) \cite{BHvK02,HJP17} is very efficient to describe halo nuclei within nuclear-reaction codes \cite{CPH18,YC18,HC21,Cap22}.
In this talk, we illustrate this and extend the idea to two new applications.
First, we show that Halo-EFT is ideal to analyse Coulomb-breakup reactions within a Bayesian approach \cite{BSC26}.
We particularise this to the breakup of $^{19}$C on $^{208}$Pb at $67A$\,MeV measured at RIKEN \cite{Nak99} and infer very precise estimates for the one-neutron separation energy $S_{\rm n}$ and Asymptotic Normalisation Constant (ANC) of that nucleus.

Second, we expand Halo-EFT to include core excitation \cite{KC25} and use that model to reanalyse the breakup of $^{11}$Be on $^{208}$Pb at $69A$\,MeV measured also at RIKEN \cite{Fuk04}.
That new analysis confirms that the reaction is purely peripheral, in the sense that it probes only the tail of the projectile wave function, viz. its $S_{\rm n}$ and ANC \cite{CN07}.
In particular, it is by no means sensitive to the so-called spectroscopic factor (SF).
Consequently, these values should not be inferred from Coulomb-breakup data \cite{KC26}.

After a brief description of the theoretical framework of this study, including a short description of Halo-EFT, we summarise in \Sec{19C} our Bayesian analysis of the Coulomb breakup of $^{19}$C \cite{BSC26}.
Then, in \Sec{11Be}, we present our study of the Coulomb breakup of $^{11}$Be performed in a coupled-channel description of the projectile \cite{KC26}.
We conclude with a brief outlook in \Sec{conclusion}.

\section{Theoretical Framework}

\subsection{Few-Body Model of Breakup Reactions}
To describe reactions involving one-neutron halo nuclei, we consider a usual three-body model \cite{BC12}.
The projectile $P$ is described as a core $c$ assumed in its ground state, to which a neutron n is loosely bound.
That two-body system is described by the effective Hamiltonian
\beq
H_0=T_r+V_{c\rm n}(\ve{r}),
\eeqn{e1}
where $\ve{r}$ is the $c$-n relative coordinate and $V_{c\rm n}$ is an effective potential fitted to reproduce what is known about the projectile's structure, such as its binding energy and excited states.

The target $T$ is seen as a structureless particle and its interaction with the projectile constituents is simulated by optical potentials chosen in the literature.
Within this model, the calculation of breakup reactions reduces to solving the three-body scattering problem
\beq
\left[T_R+H_0+V_{cT}+V_{{\rm n}T}\right]\Psi(\ve{r},\ve{R})={\cal E}\ \Psi(\ve{r},\ve{R}),
\eeqn{e2}
where $\ve{R}$ is the coordinate of the projectile centre of mass relative to the target, $V_{cT}$ and $V_{{\rm n}T}$ are the optical potentials for the $c$-$T$ and n-$T$ interactions, respectively, and ${\cal E}$ is the total energy in the centre-of-mass rest frame.
Equation \eq{e2} is solved with the condition that the projectile, initially in its ground state $\varphi_0$, is impinging on the target:
\beq
\Psi(\ve{r},\ve{R})\flim{Z}{-\infty}e^{iKZ+\cdots}\ \varphi_0(\ve{r}).
\eeqn{e3}
The direction of the initial $P$-$T$ relative momentum $\ve{K}$ is chosen as the $Z$ axis and its norm is related to ${\cal E}$ and the $c$-n ground state energy $E_0$: $\hbar^2K^2/2\mu_{PT}={\cal E}+E_0$, with $\mu_{PT}$ the $P$-$T$ reduced mass.

Various numerical methods have been developed to solve \Eq{e2}, see Ref.\,\cite{BC12} for a review.
We use the Coulomb-Corrected Eikonal approximation (CCE) \cite{CBS08} to study the Coulomb breakup of $^{19}$C in \Sec{19C}.
Our analysis of the Coulomb breakup of $^{11}$Be in \Sec{11Be} is performed at the first order of the perturbation theory \cite{AW75}.

\subsection{Halo-EFT}\label{haloeft}
As mentioned above, we consider an effective $c$-n interaction to describe the projectile's structure.
Specially, we consider a Halo-EFT description \cite{BHvK02,HJP17}.
Halo-EFT is based on the clear separation of scales that exists in halo nuclei: the ratio of the compact radius of the core to the extended size of the halo provides a small parameter upon which the $c$-n Hamiltonian can be expanded.

Neglecting the short-range physics, viz. the internal structure of the core, we describe the $c$-n interaction by a contact force and its derivatives.
For practical purposes, the contact interaction is regularised by a Gaussian of width $\sigma$, which corresponds to the range of the physics neglected in the EFT.
The Low-Energy Constants (LECs) of that interaction are fitted in different $c$-n partial waves $lj$.
At next-to-leading order, we parametrise it in the $s$ and $p$ waves as
\beq
V_{lj}(r)=V_0^{lj}\ e^{-\frac{r^2}{2\sigma^2}}+V_2^{lj}\ r^2e^{-\frac{r^2}{2\sigma^2}},
\eeqn{e4}
where $V_0^{lj}$ and $V_2^{lj}$ are the LECs fitted to what is known about the nucleus.
Typically for a one-neutron halo nucleus with a $\frac{1}{2}^+$ ground state, such as $^{19}$C and $^{11}$Be, they are adjusted in the $s1/2$ partial wave to reproduce the one-neutron separation energy $S_{\rm n}$ and the ANC ${\cal C}_{s1/2}$ of the nucleus' bound state \cite{CPH18}.
In higher partial waves, it can be adjusted to reproduce known excited states, $c$-n effective-range parameters, or single-particle resonances \cite{CPH18}.

Halo-EFT has been shown to be very efficient in the description of various reactions involving halo nuclei, such as breakup \cite{CPH18}, transfer \cite{YC18}, and knockout \cite{HC21}, see Ref.\,\cite{Cap22} for a review.
We use it in two ways.
First, within a Bayesian analysis of the Coulomb breakup of $^{19}$C to constrain the LECs upon experimental cross sections. 
Second, in an extension of Halo-EFT that includes the excitation of the core to test the sensitivity of Coulomb breakup to SFs. 

\section{Bayesian Analysis of the Coulomb Breakup of $^{19}$C}\label{19C}

\subsection{Bayesian Approach in a Nutshell}\label{bayesian}
The Bayesian method offers a statistical way to constrain theoretical models from experimental data.
Bayesian approaches \cite{gelman1995, sivia2006} are increasingly in use in nuclear physics.
They are already applied in the context of effective field theories, most notably in chiral EFT \cite{furnstahl2015} but also to analyse transfer \cite{lovell2017} and knockout reactions \cite{hebborn2023}.
The method is based on Bayes theorem, which relates different conditional probabilities
\beq
\mathrm{pr}(\ve{a}|D)=\frac{\mathrm{pr}(D|\ve{a})\,\mathrm{pr}(\ve{a})}{\mathrm{pr}(D)}.
\eeqn{e10}
In this form, it provides the posterior distribution $\mathrm{pr}(\ve{a}|D)$, viz. the probability distribution of parameters $\ve{a}$, knowing the data $D=\left \{\sigma_j^{\mathrm{exp}}\pm\Delta_j\right\}$ and assuming a prior distribution on the model parameters $\mathrm{pr}(\ve{a})$.
The confrontation of the theoretical predictions $\left\{\sigma^{\mathrm{th}}_j\right\}$ with the data is done through the likelihood 
\beq
\mathrm{pr}(D|\ve{a})=\prod_j\frac{\exp\left[-\frac{\left(\sigma^{\mathrm{th}}_j-\sigma^{\mathrm{exp}}_j\right)^2}{2\Delta_j^2}\right]}{\sqrt{2\pi}\Delta_j}.
\eeqn{e11}


\subsection{Coulomb Breakup of $^{19}$C on Pb at $67A$\,MeV}
We reanalyse the Coulomb breakup of ${19}$C, which has been measured on $^{208}$Pb at RIKEN with a beam of $67A$\,MeV \cite{Nak99}.
To sample as large a model space as possible, we consider very broad uniform priors: $V_0^{s_{1/2}}\in[-100,+100]$\,MeV \& $V_2^{s_{1/2}}\in[-100,+150]$\,MeV\,fm$^{-2}$.
We run $10^6$ reaction calculations to get statistical confidence \cite{BSC26}.

\begin{figure}[ht]
\includegraphics[width=0.5\linewidth]{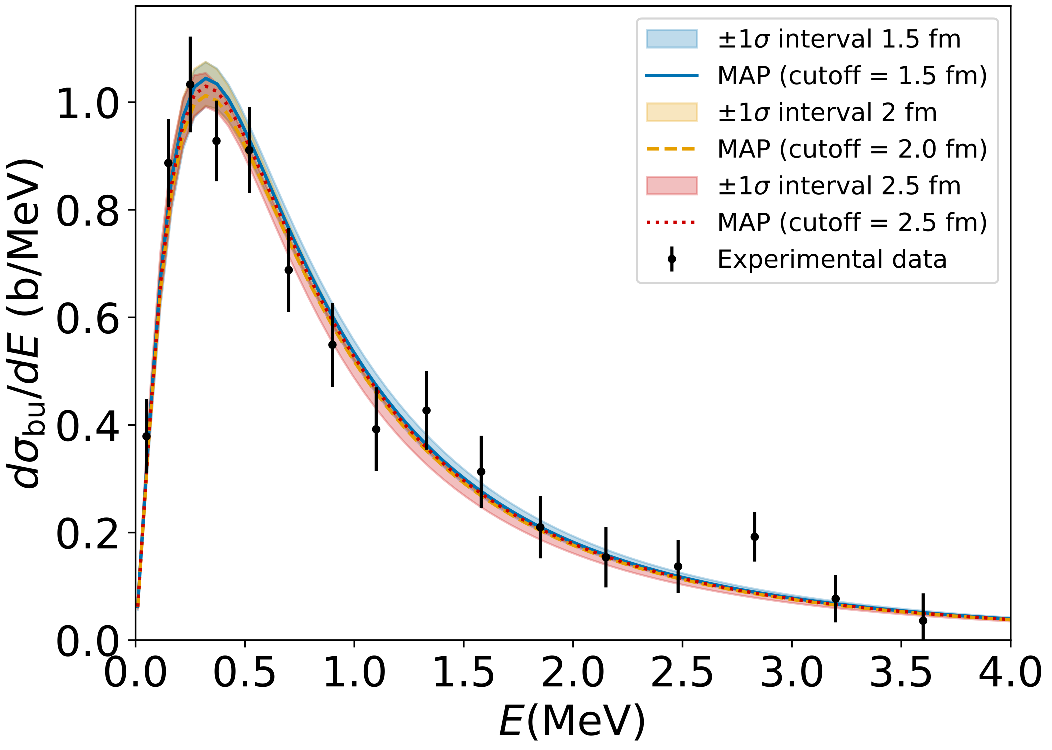}
\includegraphics[width=0.5\linewidth]{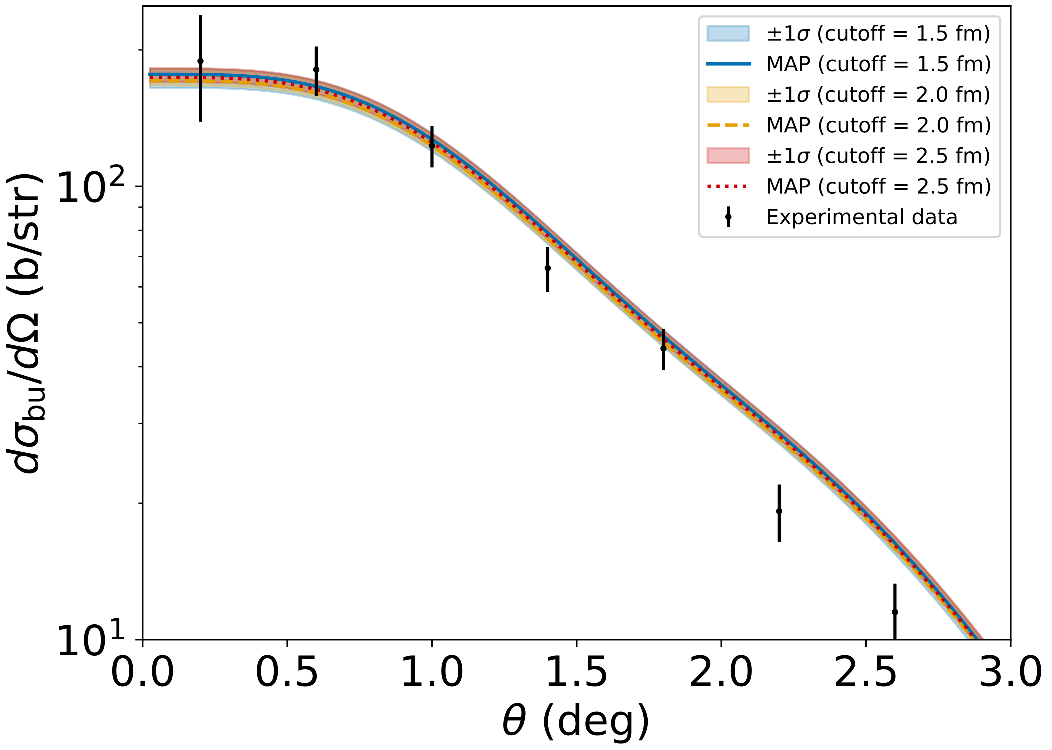}
\caption{Bayesian analysis of the Coulomb breakup of $^{19}$C on Pb at $67A$\,MeV.
(Left) energy distribution used to compute the likelihood \eq{e11}.
(Right) angular distribution used to check the quality of the fit.
In both cases, the agreement with the measurements of Ref.\,\cite{Nak99} is excellent.
}\label{fig1}
\end{figure}

The breakup cross sections obtained through this analysis are shown in \Fig{fig1}.
The left panel corresponds to the energy distribution, viz. the breakup cross section expressed as a function of the $^{18}$C-n relative energy $E$ after dissociation.
The right panel shows the angular distribution, viz. the breakup cross section as a function of the scattering angle $\theta$ of the $^{18}$C-n centre of mass.
In both cases, our results are confronted to the measurements of Ref.\,\cite{Nak99}.

We vary the LECs of the $^{18}$C-n Halo-EFT Hamiltonian \eq{e4} in the $s1/2$ partial wave of the ground state; in the other partial waves, the $^{18}$C-n continuum is described by plane waves.
From the broad prior discussed in \Sec{bayesian}, we obtain a posterior for the LECs using the likelihood \eq{e11} computed on the sole energy distribution.
The solid lines in \Fig{fig1} correspond to the maximum a posteriori (MAP), whereas the light-coloured bands show the 1-$\sigma$ confidence intervals.
To test the sensitivity of the reaction to short-range physics, we perform the calculations with three different regulators: $\sigma=1.5$\,fm (blue), 2\,fm (yellow), and 2.5\,fm (red).
All three values provide nearly identical results, confirming that the reaction is not sensitive to the internal part of the wave function, and hence that it is peripheral, in the sense that it probes only the tail of the projectile wave function \cite{CN07}.
These results are in excellent agreement not only with the energy distribution (left panel of \Fig{fig1}), but also with the angular distribution (right panel of \Fig{fig1}), although these data have not been used to constrain the LECs.
This near-perfect agreement validates the present approach.

\begin{figure}[h]
\vspace{-4mm}
\begin{minipage}[b]{0.39\linewidth}
\includegraphics[width=\linewidth]{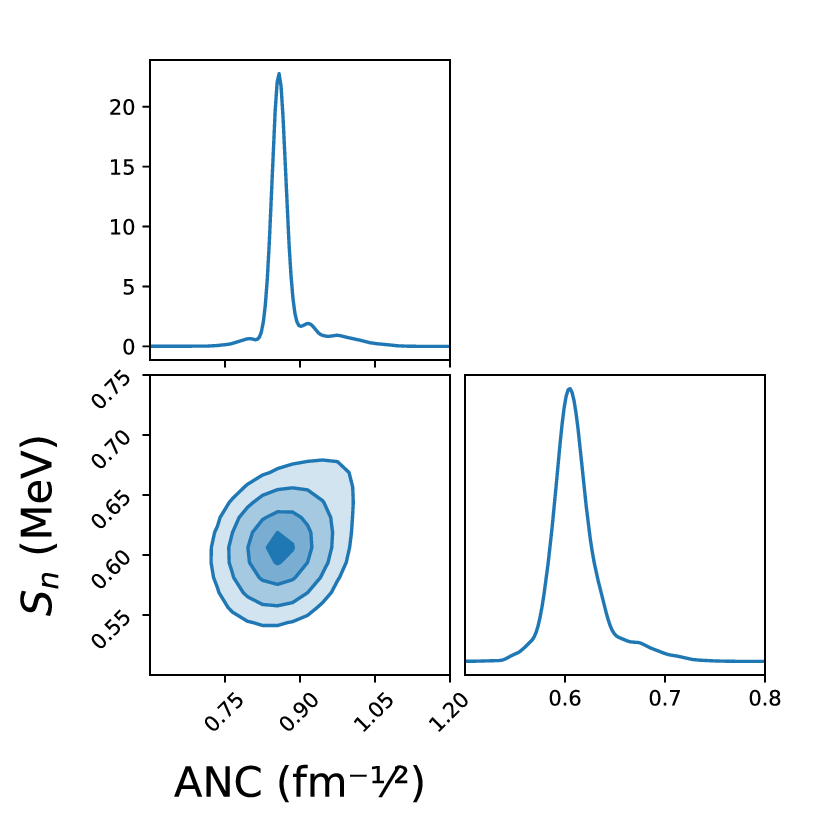}
\end{minipage}
\begin{minipage}[b]{0.60\linewidth}
\caption{Corner plot of the physical structure observables $S_n$ and ${\cal C}_{s1/2}$ of $^{19}$C inferred from the Bayesian analysis of the Coulomb-breakup data of $^{19}$C of Ref.\,\cite{Nak99}.
}\label{fig2}
\end{minipage}
\end{figure}

From the posterior on the LECs, we can then deduce a posterior on the physical observables of $^{19}$C, viz. its $S_{\rm n}$ and ANC ${\cal C}_{s1/2}$.
The corresponding corner plot is shown in \Fig{fig2}; all cutoffs of the Halo-EFT provide similar results.
We first note that for both structure observables, the posterior distribution is very narrow.
Their value is therefore well constrained by the data:
\bit
\item $S_{\rm n}=0.60\pm0.02$\,MeV, improving the tabulated value $S_{\rm n}=0.17\pm0.37$\,MeV \cite{Masses21};
\item ${\cal C}_{1/2^+}=0.85\pm0.02$\,fm$^{-1/2}$.
\eit
Note however, that these small uncertainties include only the experimental errors, and do not consider the sensitivity to model inputs, such as the choice of the optical potentials $V_{cT}$ and $V_{{\rm n}T}$, and higher orders in the Halo-EFT expansion.

\newpage
\section{Unobservability of Spectroscopic Factors in Breakup Reaction}\label{11Be}

\subsection{Spectroscopic Factors}

So far, we have assumed a simple single-particle description of halo nuclei, with an inert core assumed to be in its ground state, to which a neutron is loosely bound.
However reality is more complex.
In particular for the one-neutron halo nucleus $^{11}$Be, its ground state wave function could include components in which the $^{10}$Be core is in its first $2^+$ excited state:
\beq
    \Psi^{1/2^+}\left(^{11}{\rm Be}\right)= \varphi_{1s_{1/2}}\otimes\phi^{0^+}\left(^{10}{\rm Be}\right) + \varphi_{0d_{5/2}}\otimes\phi^{2^+}\left(^{10}{\rm Be}\right)+ \varphi_{0d_{3/2}}\otimes\phi^{2^+}\left(^{10}{\rm Be}\right)
\eeqn{e100}
In that case, the overlap wave function $\varphi_{1s_{1/2}}$ will no longer be normed to unity but it would correspond to a SF ${\cal S}_{1s_{1/2}\otimes 0^+}= \|\varphi_{1s_{1/2}}\|^2\ <1$.

Assuming the former channel to be dominant in the reaction process, and taking into account the linearity of the Schr\"odinger \Eq{e2}, ${\cal S}_{1s1/2\otimes 0^+}$ is usually inferred from experiment by \cite{HT03}
\beq
   {\cal S}_{1s_{1/2}\otimes 0^+} = \frac{\sigma^{\rm exp}}{\sigma^{\rm th}},
\eeqn{e101}
where $\sigma^{\rm exp}$ is the experimental cross section and $\sigma^{\rm th}$ is the theoretical prediction computed assuming a single-particle structure normed to unity.

We test the assumption \eq{e101} by performing breakup calculations with a description of halo nuclei that includes core excitation, and hence that leads to the multi-channel wave function \eq{e100}.
If \Eq{e101} were valid, we should obtain theoretical cross sections that scale with ${\cal S}_{1s_{1/2}\otimes 0^+}$.
Here, we treat the reaction at the first order of the perturbation theory \cite{AW75}.

\subsection{Including Core Excitation in the Description of Halo Nuclei}

To include the core excitation, we consider the effective $c$-n Hamiltonian
\beq
 H(\ve{r},\xi)= -\frac{\hbar^2}{2\mu}\Delta + V_{c\rm n}({\ve{r}},\xi) + h_{c}({\xi}),
\eeqn{e102}
where $h_{c}$ is the intrinsic core Hamiltonian with eigenstates $\phi_{M_{c}}^{I_{c}^{\pi_c}}$.
We follow Nunes \etal \cite{NTJ96} and consider a particle-rotor model \emph{à la} Bohr and Mottelson \cite{BM69}.
The $c$-n effective potential now depends on the intrinsic coordinates of the core $\xi$
\beq
    V_{cn}(\ve{r},\xi)= V(r) + \beta\, \sigma\, Y_{2}^{0}(\hat{r}') \frac{d}{d\sigma}V(r),
\eeqn{e103}
where $\hat{r}'$ is the solid angle of the $c$-n coordinate expressed in the intrinsic reference frame of the deformed core, which depends on $\xi$.
We choose as potential $V$ the Gaussian interaction \eq{e4}. 
In the usual approach, $\beta$ corresponds to the quadrupole deformation of the core.
We see it merely as a parameter that couples the different configurations.
By varying it, we can move probability amplitude from the main $1s_{1/2}\otimes 0^+$ channel towards the $0d_j\otimes 2^+$ ones to change ${\cal S}_{1s_{1/2}\otimes 0^+}$.

To obtain the eigenstates of $H$ \eq{e102}, we expand the wave function onto the core eigenstates
\beq
    \Psi^{J^\pi M}(\ve{r},\xi)=\sum_{\alpha} i^\ell \frac{u_{\alpha}(r)}{r} [\mathcal{Y}_{\ell j}(\hat{r}) \otimes \phi^{I_{c}^{\pi_c}}(\xi)]^{J M}.
\eeqn{e104}
This leads to a set of coupled equations
\beq
\left\{\frac{\hbar^2}{2\mu}\left[-\frac{d^2}{dr^2}+\frac{\ell(\ell+1)}{r^2}\right]+V_{\alpha \alpha}(r) + \epsilon_{\alpha}- E\right\} u_{\alpha}(r) =  -\sum_{\alpha'\neq \alpha}V_{\alpha \alpha'}(r)u_{\alpha'}(r),
\eeq
which we solve within the R-Matrix method on a Lagrange mesh \cite{Bay15,KC25}.

We fit the LECs of the deformed $c$-n interaction in the $\frac{1}{2}^+$ wave to reproduce $S_{\rm n}(^{11}{\rm Be})=0.50$\,MeV and the ANC ${\cal C}_{s_{1/2}}=0.786$\,fm$^{-1/2}$ predicted \emph{ab initio} in Ref.\,\cite{Cal16}.

\subsection{Re-Analysis of the Coulomb Breakup of $^{11}$Be on Pb at $69A$\,MeV}
\begin{figure}
\bce
\includegraphics[width=0.49\linewidth]{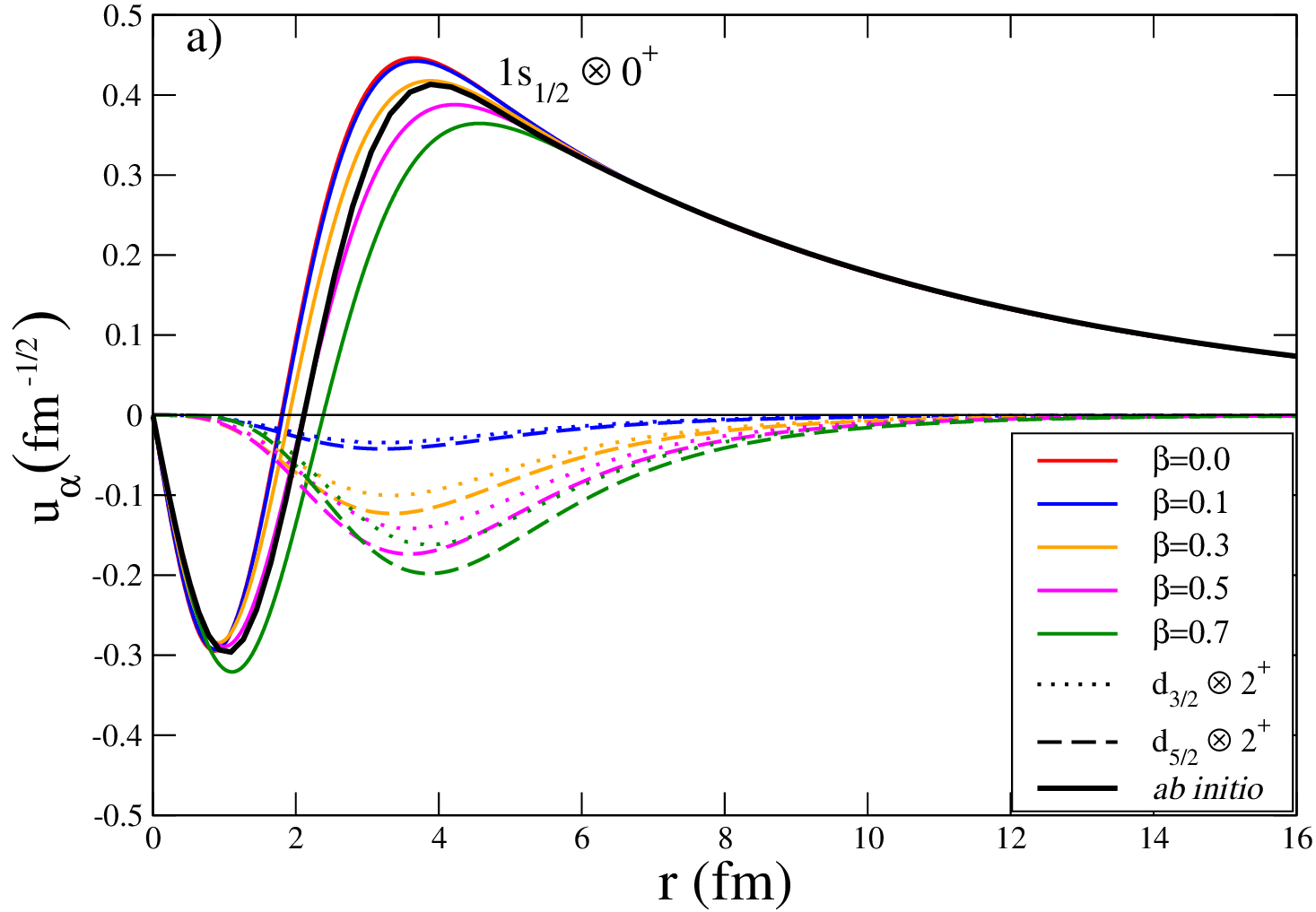}
\includegraphics[width=0.49\linewidth]{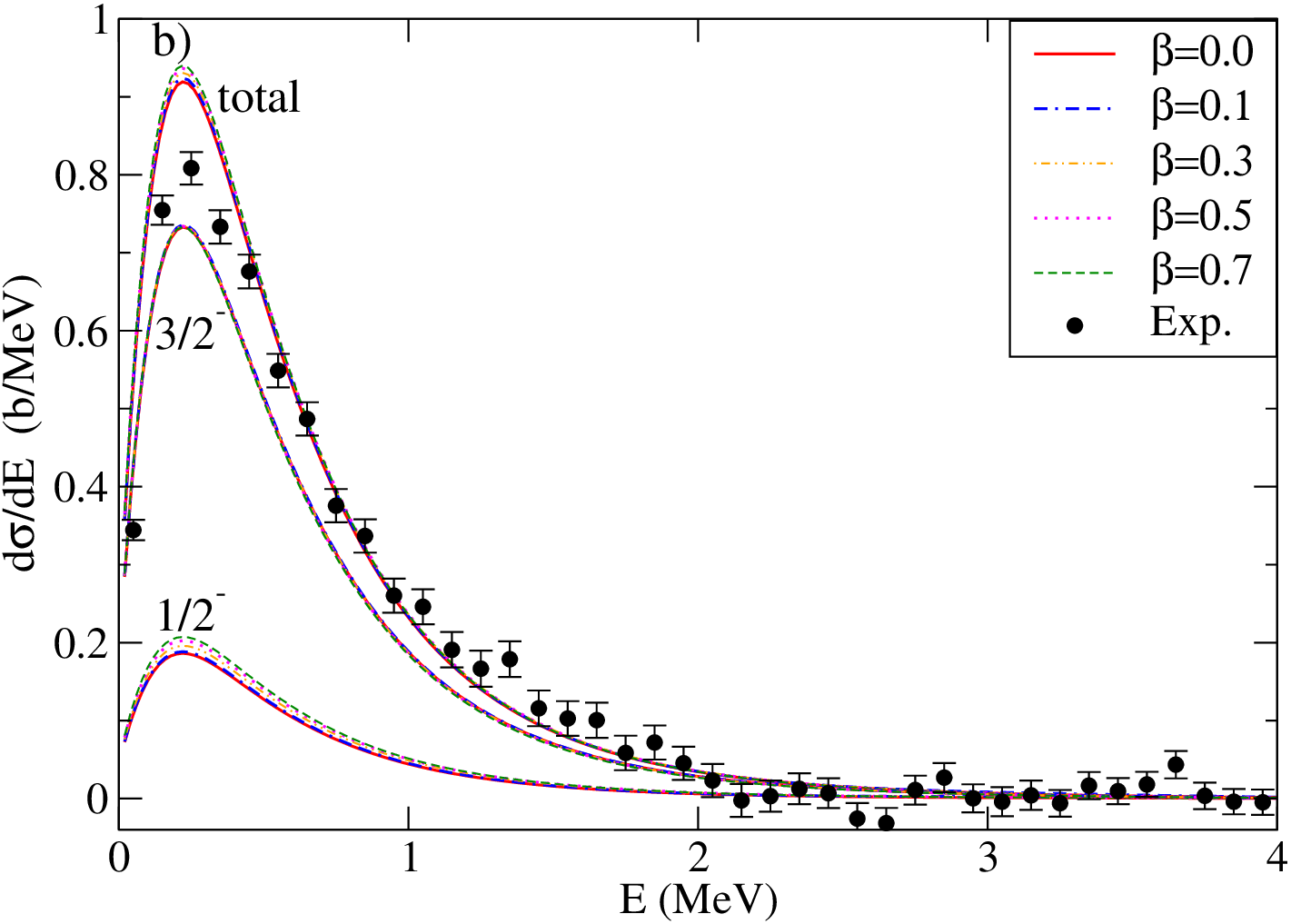}
\ece
\caption{Re-analysis of the Coulomb breakup of $^{11}$Be on Pb at $69A$\,MeV \cite{Fuk04}.
(a) Radial wave functions of $^{11}$Be $\frac{1}{2}^+$ ground state in the coupled-channel model including both the $0^+$ ground and first $2^+$ excited states of the $^{10}$Be core. The different colours correspond to different coupling strength $\beta$ as shown in the legend.
(b) Resulting Coulomb-breakup cross section obtained at the first order of the perturbation theory.
Figures are from Ref.\,\cite{KC26}.
}\label{fig3}
\end{figure}

The radial wave functions obtained from this description of $^{11}$Be are shown in \Fig{fig3}(a) for $\beta=0$--0.7.
By construction, all the main configurations $1s_{1/2}\otimes 0^+$ exhibit the same tail beyond $r\approx6$\,fm.
However, the short-range part of the wave function varies significantly with $\beta$.
When the coupling strength increases, probability strength moves from the $1s_{1/2}\otimes 0^+$ channel (solid lines) towards the $0d_j\otimes 2^+$ ones (dotted and dashed lines) \cite{KC26}.
For $\beta=0.3$--0.5, we obtain a good agreement with the \emph{ab initio} prediction of Ref.\,\cite{Cal16} (thick black line).

We use these different descriptions of $^{11}$Be within a model of Coulomb breakup based on the first order of the perturbation theory \cite{AW75}.
The resulting cross sections for the breakup of $^{11}$Be on $^{208}$Pb at $69A$\,MeV are shown as a function of the $^{10}$Be-n relative energy $E$ after dissociation in \Fig{fig3}(b).
The main result is that all bound-state wave functions shown in \Fig{fig3}(a) lead to the same breakup cross section.
Even though the spectroscopic factor ${\cal S}_{1s_{1/2}\otimes 0^+}$ varies from 1 ($\beta=0$, red solid line) to 0.8 ($\beta=0.7$; green dashed line), the corresponding reaction observable remains unchanged.
This confirms that Coulomb breakup is actually not sensitive to the short range of the projectile wave function.
As already shown in Ref.\,\cite{CN07} and in \Sec{19C}, the reaction is sensitive mostly to $S_{\rm n}$ and to the ANC.
Using a multi-channel description of the projectile, the present analysis demonstrates that Coulomb breakup is insensitive to SFs \cite{KC26}.
The good agreement with the RIKEN data of \cite{Fuk04} validates our approach.

\section{Conclusion}\label{conclusion}
Halo-EFT \cite{BHvK02,HJP17} is very efficient to describe halo nuclei within accurate models of reactions \cite{CPH18,YC18,HC21,Cap22}.
In this talk, we have illustrated its use in further applications.

First, we consider Halo-EFT to analyse reaction data within a Bayesian approach.
In our reanalysis of the Coulomb breakup of $^{19}$C, we have shown that a narrow posterior distribution of the Halo-EFT LECs can be obtained.
The excellent agreement obtained with experiment validates the MAP values hence obtained.
This then enables us to infer precise values for structure observables of $^{19}$C: its binding energy and ANC of the ground state \cite{BSC26}.

Extending Halo-EFT to include core excitation, we have shown that this excitation, and, consequently, the so-called spectroscopic factors, have no influence on the breakup cross section.
Even a change of 20\% of the SF of $^{11}$Be does not affect the reaction observables.
Spectroscopic factors should therefore not be inferred from Coulomb-breakup measurement.

In the future, we plan to extend our Bayesian analysis to include other model inputs such as the optical potentials simulating the $P$-$T$ interaction and higher orders of the Halo-EFT.
We plan also to improve our model of breakup reactions with core excitation to account for the nuclear part of the $P$-$T$ interaction.
This will also enable us to study other reactions, such as nuclear-dominated breakup and knockout.

	\section*{Acknowledgements}
This work was supported by the Deutsche Forschungsgemeinschaft (DFG, German Research Foundation) through Project-ID 279384907 – SFB 1245 and the Cluster of Excellence “Precision Physics, Fundamental Interactions and Structure of Matter” (PRISMA++ EXC 2118/2, Project ID No. 390831469), and by the U.S.~Department of Energy under contract No.~DE-FG02-93ER40756.


\begin{thebibliography}{10}

\bibitem{Tan85b}
I.~Tanihata, H.~Hamagaki, O.~Hashimoto, S.~Nagamiya, Y.~Shida, N.~Yoshikawa,
  O.~Yamakawa, K.~Sugimoto, T.~Kobayashi, D.E. Greiner, N.~Takahashi, and
  Y.~Nojiri.
\newblock Measurements of interaction cross sections and radii of He isotopes.
\newblock {Phys. Lett. B} {\bf 160}, 380 (1985).

\bibitem{Tan96}
I.~Tanihata.
\newblock Neutron halo nuclei.
\newblock {J. Phys. G} {\bf 22},157 (1996).

\bibitem{BHvK02}
C.A. Bertulani, H.-W. Hammer, and U.~{van Kolck}.
\newblock Effective field theory for halo nuclei: shallow $p$-wave states.
\newblock {Nucl. Phys. A} {\bf 712.}, 37 (2002).

\bibitem{HJP17}
H.-W.~Hammer, C.~Ji, and D.~R. Phillips.
\newblock Effective field theory description of halo nuclei.
\newblock {J. Phys. G} {\bf 44}, 103002 (2017).

\bibitem{CPH18}
P.~Capel, D.~R. Phillips, and H.-W. Hammer.
\newblock Dissecting reaction calculations using halo effective field theory
  and \emph{ab initio} input.
\newblock {Phys. Rev. C} {\bf 98}, 034610 (2018).

\bibitem{YC18}
J.~Yang and P.~Capel.
\newblock Systematic analysis of the peripherality of the
  $^{10}\mathrm{Be}(d,p)^{11}\mathrm{Be}$ transfer reaction and extraction of
  the asymptotic normalization coefficient of $^{11}\mathrm{Be}$ bound states.
\newblock {Phys. Rev. C} {\bf 98}, 054602 (2018).

\bibitem{HC21}
C.~Hebborn and P.~Capel.
\newblock Halo effective field theory analysis of one-neutron knockout
  reactions of $^{11}\mathrm{Be}$ and $^{15}\mathrm{C}$.
\newblock {Phys. Rev. C} {\bf 104}, 024616 (2021).

\bibitem{Cap22}
Pierre Capel.
\newblock Combining {Halo-EFT} descriptions of nuclei and precise models of
  nuclear reactions.
\newblock {Few-Body Syst.} {\bf 63}, 14 (2022).

\bibitem{BSC26}
Q.~Bozet, I.~Svensson, and P.~Capel.
\newblock Bayesian analysis of the {Coulomb} breakup of $^{19}${C} (2026).
\newblock (In preparation).

\bibitem{Nak99}
T.~Nakamura, N.~Fukuda, T.~Kobayashi, N.~Aoi, H.~Iwasaki, T.~Kubo, A.~Mengoni,
  M.~Notani, H.~Otsu, H.~Sakurai, S.~Shimoura, T.~Teranishi, Y.~X. Watanabe,
  K.~Yoneda, and M.~Ishihara.
\newblock Coulomb dissociation of {$^{19}$C} and its halo structure.
\newblock {Phys. Rev. Lett.} {\bf 83}, 1112 (1999).

\bibitem{KC25}
L.-P. Kubushishi and P.~Capel.
\newblock Exploring core excitation in halo nuclei using halo effective field
  theory: an application to the bound states of $^{11}${Be}.
  \newblock  arXiv:2507.13585 (2025).

\bibitem{Fuk04}
N.~Fukuda, T.~Nakamura, N.~Aoi, N.~Imai, M.~Ishihara, T.~Kobayashi, H.~Iwasaki,
  T.~Kubo, A.~Mengoni, M.~Notani, H.~Otsu, H.~Sakurai, S.~Shimoura,
  T.~Teranishi, Y.~X. Watanabe, and K.~Yoneda.
\newblock Coulomb and nuclear breakup of a halo nucleus $^{11}\mathrm{Be}$.
\newblock {Phys. Rev. C} {\bf 70}, 054606 (2004).

\bibitem{CN07}
P.~Capel and F.~M. Nunes.
\newblock Peripherality of breakup reactions.
\newblock {Phys. Rev. C} {\bf 75}, 054609 (2007).

\bibitem{KC26}
L.-P. Kubushishi and P.~Capel.
\newblock Insensitivity of the Coulomb breakup of halo nuclei to spectroscopic
  factors.
\newblock {Phys. Lett. B} {\bf 879}, 140694 (2026).

\bibitem{BC12}
D.~Baye and P.~Capel.
\newblock Breakup reaction models for two- and three-cluster projectiles.
\newblock {Lecture Notes in Physics} {\bf 848}, 121 (2012).
\newblock {(Ed. C. Beck)}.

\bibitem{CBS08}
P.~Capel, D.~Baye, and Y.~Suzuki.
\newblock Coulomb-corrected eikonal description of the breakup of halo nuclei.
\newblock {Phys. Rev. C} {\bf 78}, 054602 (2008).

\bibitem{AW75}
K.~Alder and A.~Winther.
\newblock {\em Electromagnetic Excitation: Theory of Coulomb Excitation with
  Heavy Ions}.
\newblock North-Holland Publishing Company, 1975.

\bibitem{gelman1995}
A.~Gelman, J.B.~Carlin, H.S.~Stern, and D.B.~Rubin.
\newblock {\em Bayesian data analysis}.
\newblock Chapman and Hall/CRC, 1995.

\bibitem{sivia2006}
D.~Sivia and J.~Skilling.
\newblock {\em Data analysis: a Bayesian tutorial}.
\newblock OUP Oxford, 2006.

\bibitem{furnstahl2015}
R.J.~Furnstahl, D.R.~Phillips, and S.~Wesolowski.
\newblock A recipe for {EFT} uncertainty quantification in nuclear physics.
\newblock {J. Phys. G} {\bf 42}, 034028 (2015).

\bibitem{lovell2017}
A.E.~Lovell, F.M.~Nunes, J.~Sarich, and S.M.~Wild.
\newblock Uncertainty quantification for optical model parameters.
\newblock {Phys. Rev. C} {\bf 95}, 024611 (2017).

\bibitem{hebborn2023}
C.~Hebborn, T.R.~Whitehead, A.E.~Lovell, and F.M.~Nunes.
\newblock Quantifying uncertainties due to optical potentials in one-neutron
  knockout reactions.
\newblock {Phys. Rev. C} {\bf 108}, 014601 (2023).

\bibitem{Masses21}
F.G. Kondev, M.~Wang, W.J. Huang, S.~Naimi, and G.~Audi.
\newblock The {NUBASE}2020 evaluation of nuclear physics properties.
\newblock {Chinese Phys. C} {\bf 45}, 030001 (2021).

\bibitem{HT03}
P.G. Hansen and J.A. Tostevin.
\newblock {Direct Reactions with Exotic Nuclei}.
\newblock {Ann. Rev. Nucl. Part. Sc.} {\bf 53}, 219 (2003).

\bibitem{NTJ96}
F.M. Nunes, I.J. Thompson, and R.C. Johnson.
\newblock Core excitation in one neutron halo systems.
\newblock {Nucl. Phys. A} {\bf 596}, 171 (1996).

\bibitem{BM69}
A.~Bohr and B.~Mottelson.
\newblock {\em Nuclear Structure}.
\newblock W. A. Benjamin, New York, 1969.

\bibitem{Bay15}
D.~Baye.
\newblock {The Lagrange-mesh method}.
\newblock {Phys. Rep.} {\bf 565}, 1 (2015).

\bibitem{Cal16}
A.~Calci, P.~Navr\'atil, R.~Roth, J.~Dohet-Eraly, S.~Quaglioni, and G.~Hupin.
\newblock Can ab initio theory explain the phenomenon of parity inversion in
  $^{11}\mathrm{Be}$?
\newblock {Phys. Rev. Lett.} {\bf 117}, 242501 (2016).

\end{thebibliography}

\providecommand{\noopsort}[1]{}\providecommand{\singleletter}[1]{#1}%

\end{document}